\documentclass[showpacs,superscriptaddress,two column,pra,floatfix,nofootinbib]{revtex4-1}
\usepackage{amsmath}
\usepackage{amssymb}
\usepackage{amsthm}
\usepackage{latexsym}
\usepackage{array}
\usepackage{amsfonts}
\usepackage{mathrsfs}
\usepackage{color}
\usepackage{hyperref}
\usepackage{graphicx}
\usepackage{multirow}
\def\be{\begin{equation}}
\def\ee{\end{equation}}
\def\bea{\begin{eqnarray}}
\def\eea{\end{eqnarray}}
\def\ben{\begin{equation*}}
\def\een{\end{equation*}}
\def\bean{\begin{eqnarray*}}
\def\eean{\end{eqnarray*}}
\def\bma{\begin{mathletters}}
\def\ema{\end{mathletters}}
\def\bi{\begin{itemize}}
\def\ei{\end{itemize}}

\newtheorem{theorem}{Theorem}

\newtheorem{observation}[theorem]{Observation}

\begin{document}
\title{All-versus-nothing violation of local realism from the Hardy paradox under no-signaling}

%&&&&&&&&&&&&&&&&&&&&&&&&&&&&&&&&&&&&&&&&&&&&&&&&
\author{Some Sankar Bhattacharya}
\email{somesankar@gmail.com}\affiliation{Physics \& Applied Mathematics Unit, Indian Statistical Institute, 203 B.T. Road, Kolkata- 700108, India}
\author{Arup Roy}
\email{arup145.roy@gmail.com}\affiliation{Physics \& Applied Mathematics Unit, Indian Statistical Institute, 203 B.T. Road, Kolkata- 700108, India}
\author{Amit Mukherjee}
\email{amitisiphys@gmail.com}\affiliation{Physics \& Applied Mathematics Unit, Indian Statistical Institute, 203 B.T. Road, Kolkata- 700108, India}
\author{Ramij Rahaman}
\email{ramijrahaman@gmail.com}\affiliation{Department of Mathematics, University of Allahabad, Allahabad 211002, U.P., India}

\begin{abstract}
Hardy's is one of the simplest arguments concerning non-locality. Recently Chen {\it et. al.} have proposed a more generalized Hardy-like argument and have shown that the probability of success increases with local system's dimension. Here we study the same in a minimally constrained theory, namely the generalized no-signaling theory(GNST). We find that not only the probability of success of this argument increases with local system dimension in GNST, it also takes a very simple functional form. %The success probability also provides an {\em All-Versus-Nothing} type of violation of local realism asymptotically.
\end{abstract}

\maketitle

\section{Introduction}
In 1964, J.S. Bell, proved that one can find measurement correlations for a composite quantum system which cannot be described by local-realistic theory(LRT) \cite{Bell64}. Though the work of Bell is pioneer, the approach proof was not much impressive due to its statistical nature. Bell's inequalities \cite{CHSH69}, in fact, are statistical predictions about some sets of measurements which can be made on local subsystems far separated from each other. The violation of such inequality implies that the statistical description can not be reproduced by local hidden variables.

Greenberger, Horne and Zeilinger \cite{GHZ89} found a way to show more immediately, without inequalities, the result or prediction of quantum mechanics are inconsistent with the assumption of Einstein, Podolsky and Rosen {\em i.e} locality and reality \cite{EPR35}. Unlike Bell, this proof involves only one event and not the statistics of many events. In 1992, Hardy \cite{Har92}, gave a relatively simpler All-or-none type proof of this no-go theorem for local hidden variables without using any statistical inequalities, via some logical contradiction, in the same spirit of GHZ \cite{ghz,GHZ89}. Hardy's non-locality argument deals with two qubits with two dichotomic measurement observables on each qubit. The proof can be extended even for $n$ qubits\cite{GKS98} \cite{clif92}. The argument is also valid for more than two measurements \cite{bran} and more than two outcomes \cite{kunk,SG11,rabe,kar}. The above non-locality argument can be extended even for generalized no-signaling theory(GNST) \cite{CGKKRR10}. One can also find the opposite approach in literature, where to show that correlations originating from GNST are `more nonlocal' than quantum correlations, Tobias Fritz has considered a `stronger' version of Hardy paradox in two input, two output scenario\cite{fritz11}.   
\par Any physical theory should contain the fact that instantaneous propagation of information is impossible. This is the no-signaling principle. Non-locality obeying this principle at operational level are solely responsible for a good number of fascinating phenomena like secrecy extraction \cite{SC06}, certification of intrinsic randomness \cite{masanes} and several non-classical communication tasks.
\par The maximum success probability for Hardy's argument remains unchanged for changing system dimension. But recently Chen et.al. \cite{CCXSWK13} have formulated Hardy's non-locality argument for measurements that have more than two outcomes in a stronger way. Interestingly, the authors have shown numerically that the maximum probability of success increases with the local system's dimension. This new non-locality argument is equivalent to a violation of a tight Bell Inequality\cite{CCXSWK13}. As one might expect this argument reduces to {\em old Hardy's argument} \cite{Har92} for a special case. So it might be interesting to study this new Hardy-type argument for a minimally constrained theory, namely GNST. Recently Mansfield \cite{shane} has shown that the `Probability of witnessing Hardy Nonlocality ($PN$)' for two two-level system can be achieved with certainty under GNST. Whereas the `Paradoxical probability($PP$)' of two two-level Hardy's argument is bounded by 0.5. The $PP$ concerns the quality of a particular Hardy argument. But, $PN$ concerns the performance of a correlation regarding the demonstration of non-locality. Hence, these two concepts are motivated from different perspectives. Thus the study of $PP$ and $PN$ and their relation under GNST is worthy. This in effect can provide us with an upper bound in the paradoxical probability of this argument that is allowed by relativistic causality alone, in absence of any further constraints. This study is also important because the optimal success probability of Hardy's argument is deeply connected with the security proof of several information processing tasks \cite{Rabelo, RMPM15, RMM14}. Here we deal with these questions.
\par
We also investigate how this maximum probability of success changes with the local system's dimension and provide an analytic functional form for such a feature. We extend our work even for situations where the dimensions of spatially separated parties are not equal. The rest of the article is arranged in the following manner: section \ref{sec2} reviews the conventional and new Hardy type arguments and the results known so far, in section \ref{sec_nst} we set the stage for calculating the maximum success probability of new Hardy type paradox for higher dimensional systems in no-signaling paradigm, in section \ref{sec_res} we present our results and finally conclude in section \ref{concl}.

\section{Bi-partite Hardy paradox}
\label{sec2}
Consider a physical system consisting of two subsystems shared between two distant parties Alice and Bob. The two observers (Alice and Bob) have access to one subsystem each. Assume that Alice (Bob) can measure one of two observables, $X_0$ and $X_1$ ($Y_0$ and $Y_1$), on her (his) local subsystem. The outcomes $a(b)$ of each such Von Neumann measurement can be $1,2,...,d_X^A(d_Y^B)$. Here $d_X^A(d_Y^B)$ is the dimension of Hilbert space associated to the Alice's (Bob's) subsystem. The joint probability $P(X=a, Y=b)$ denotes the probability of getting outcome $(a,b)$ for the measurement $(X,Y)$.
\subsection{Hardy Paradox}
The pioneering non-inequality paradox regarding incompatibility of any theory with local realism, introduced by L.Hardy \cite{Har92} in 1992 is for two two-level systems. A generalized version of this argument for two multi-level systems starts with the following set of joint probability conditions:

\begin{equation}
\label{hardyOld}\begin{split}
P(X_0=1,Y_0=d_{Y_0}^B) &= q_{H}>0,\\
P(X_1=a,Y_0=d_{Y_0}^B) &= 0,~\forall a\in\{2,3,..d_{X_1}^A\},\\
P(X_0=1,Y_1=b) &= 0,~\forall b\in \{1,2,...,d_{Y_1}^B-1\},\\
P(X_1=1,Y_1=d_{Y_1}^B) &= 0.
\end{split}\end{equation}

The logical structure of the argument is as follows: for some ontic variables $\lambda\in\Omega$, the $(X_0,Y_0)$ observables can take value $(1,d_{Y_0}^B)$ which is the first condition of (\ref{hardyOld}). Let us denote the subspace span by those $\lambda$s' as $\Omega'(\subset \Omega)$. Now the second condition tells us that for all $\lambda$'s in $\Omega'$ observable $X_1$ can only take value 1, as the other possibility for $X_1$ observable is excluded. Similarly, the third condition provides us that for $Y_{1}$ observable $d_{Y_1}^B$ is the only possibility for all $\lambda\in\Omega'$. Therefore, the joint possibility for $X_{1}=1$ and $Y_{1}=d_{Y_1}^B$ should be non zero for $\lambda\in\Omega'$. But this contradicts the last condition of (\ref{hardyOld}). Therefore, the four statements of (\ref{hardyOld}) are incompatible with local-realism. But there exists quantum correlations which can reproduce all the four conditions of (\ref{hardyOld}) \cite{SG11,RWZ13}. 
\par At this point one can point out two important quantities - firstly, $q_{H}$ which is the probability of success of a stand-alone argument(\ref{hardyOld}), i.e. paradoxical probability($PP$). On the other hand, given a correlation, one can make use of two or more such arguments to demonstrate non-locality, which gives rise to $PN$. Since more than one elementary arguments are used, the complementary events(i.e. the events which are not considered in the original stand-alone argument) along with the principal events may contribute to $PN$. One can heuristically write $PP+PPC=PN$, where PPC is the probability of success contributed by the complementary events. Correlations arising from quantum systems satisfy $PP=PN$\cite{shane}. But the gap between PP and PN becomes visible when one considers post-quantum correlations. One such example is the Popescu-Rohrlic(PR) box. For PR box $PP$ corresponding to(\ref{hardyOld}) is $\frac{1}{2}$, whereas $PN=1$. Thus consideration of general post-quantum correlations reveal this curious feature of the `nonlocality without inequality'-type of argument.               

\subsection{General non-signaling theory(GNST) satisfying Hardy-type argument} In the framework of a general probabilistic theory, consider a system of two separated parties, which together satisfy all the conditions of the Hardy-type argument as given in (\ref{hardyOld}). 
\subsubsection{Positivity conditions}
For $P(X=a,Y=b)$ to be a valid probability measure it should satisfy the positivity conditions
\begin{equation}\label{pos}
P(X=a,Y=b)\geq 0~~\forall X,Y,a,b
\end{equation}

\subsubsection{Normalization conditions}
\par The probability distribution relating the outcomes for a given measurement setting should satisfy the normalization condition.
\begin{equation}\label{norm}
\sum_{a=1}^{d_X^A}\sum_{b=1}^{d_Y^B}P(X=a,Y=b)=1,
\end{equation}
$\forall X\in \{X_0,X_1\}$ and $Y\in \{Y_0,Y_1\}$.
\subsubsection{Non-signaling conditions}
For any no-signaling n-partite distribution $P(a,b,c,...|X,Y,Z,...)$ holds the fact that, each subset of parties $\{a,b,....\}$ only depends on its corresponding inputs, i.e. if we change the input of one party it does not effect the marginal probability distribution for the other spatially separated parties.
\par
For a bipartite generalized probability distribution the no-signaling conditions take the following form
\begin{eqnarray}
\sum_{b=1}^{d_{Y_0}^B}P(X=a,Y_0=b)=\sum_{b'=1}^{d_{Y_1}^B}P(X=a,Y_1=b') \nonumber \\
\forall X\in\{X_0,X_1\} \mbox{ and } a\in\{1,d_X^A\}\label{signal1}
\\
\sum_{a=1}^{d_{X_0}^A}P(X_0=a,Y=b)=\sum_{a'=1}^{d_{X_1}^A}P(X_1=a',Y=b) \nonumber \\
\forall Y\in\{Y_0,Y_1\} \mbox{ and } b\in\{1,d_Y^B\}\label{signal2}
\end{eqnarray}

What is the maximum probability of success, $P(X_0=1,Y_0=d_{Y_0}^B)$, of the Hardy-type argument (\ref{hardyOld}) under GNST for an arbitrary $d_X^A\times d_Y^B$ system, subject to the constraints given in Eq. (\ref{pos})-(\ref{signal2})? We have shown that the maximum probability of success $P(X_0=1,Y_0=d_{Y_0}^B)$ for two input, $(d_X^A,d_Y^B)$ output Hardy's test(\ref{hardyOld}) the maximum value is $\frac{1}{2}$ under GNST for all $d_X^A,d_Y^B$(Sec.\ref{res1}). Interestingly, the maximum probability of success in the bipartite Hardy-type argument under GNST is dimension independent as in the quantum case. For $2$-qubit system the maximum achievable value of Hardy's success is $q_{H}=\frac{5\sqrt{5}-11}{2}\approx0.09$ \cite{Hardy2,Jod94}. Reference \cite{SG11} proves that, for $2$-qutrit systems, maximum achievable value of Hardy's success probability is same as that of $2$-qubit system and conjectures that it will remain same for arbitrary dimension $n$. Recently, reference \cite{Rabelo} provides a proof of this conjecture. This result tells that for showing the contradiction of quantum mechanics with the \emph{local realism} higher dimensional systems give no advantage in experimental implementation of such a test. Keeping this in mind the authors in reference \cite{CCXSWK13} introduce a Hardy like argument which applies to measurements with an arbitrarily large number of outcomes. They have also showed that the success probability of this modified Hardy's paradox increases with increase in the local system's dimension.

\subsection{Relaxed Hardy Paradox}
The conditions for the new relaxed Hardy type argument \cite{CCXSWK13} are:
\begin{eqnarray}
\label{hardyNew}
&P(X_0<Y_0) = q_{RH}>0,&\nonumber\\
&P(X_1<Y_0) = 0,&\\
&P(Y_1<X_1) = 0,&\nonumber\\
&P(X_0<Y_1) = 0,&\nonumber
\end{eqnarray}
where $P(X_i<Y_j)=\sum_{a<b}P(X_i=a,Y_j=b)$. Therefore, if events $X_1<Y_0$, $Y_1<X_1$, and $X_0<Y_1$ never happen, then, in any local theory, event $X_0<Y_0$ must never happen either. However, this is not the case with quantum correlations. Both sets of conditions (\ref{hardyOld}) \& (\ref{hardyNew}) cannot be satisfied by any {\em local-realistic theory} (LRT) \cite{Har92,CCXSWK13}. One can generalize the above conditions (\ref{hardyNew}) by replacing the last zero condition with a non-zero condition $P(X_0<Y_1) =p<q_{RH}$. For $d_X^A=d_Y^B=2$, above two sets of conditions (\ref{hardyOld}) \& (\ref{hardyNew}) give us the conventional two-level Hardy paradox \cite{Har92}:
\begin{eqnarray}
\label{hardy22}
&P(X_0=1,Y_0=2) = q_{RH}=q_{H}>0,&\nonumber\\
&P(X_1=1,Y_0=2) = 0,&\\
&P(Y_1=2,X_1=1) = 0,&\nonumber\\
&P(X_0=1,Y_1=2) = 0.&\nonumber
\end{eqnarray}
 In \cite{CCXSWK13} the authors have shown that in quantum theory the success probability of relaxed Hardy test (\ref{hardyNew}) surpasses that of the conventional Hardy's test. It has also been shown that the probability of nonlocal events increases with the local system dimension and for high enough dimensions $q_{RH}$ is almost four times higher than $q_{H}$. They have also claimed that the non-locality argument proposed by them is the most natural and powerful generalization of Hardy's argument concerning higher dimension of local systems. To test such a proposal one might wonder how useful this generalized Hardy argument is, in a theory which contains minimal number of features or restrictions. In the following sections we have studied this relaxed Hardy argument in GNST, where the only restriction on theory is that it does not violate relativistic causality.
\section{Relaxed Hardy Paradox in No-signaling Theories}
\label{sec_nst}
Here we study Hardy's paradox for higher dimensional systems within the framework of generalized probabilistic theories. The only condition that we impose on the generalized probability distribution is the no-signaling condition, which all known physical theories respect.

The set of boxes which satisfy Eq.(\ref{pos})-(\ref{signal2}) can be divided into two types: local and non-local. A local box can be simulated using shared randomness only, whereas to simulate a non-local box with shared randomness, the observers must communicate. Due to the linearity of the constraints- Eq.(\ref{pos})-(\ref{signal2}), the set of all non-signaling boxes with a finite number of inputs and outputs form a polytope $\boldsymbol{P}$ with finite vertices. Convex property of such a polytope follows from the argument that a probabilistic mixture of any two boxes satisfying the linear constraints will also be a member of the polytope $\boldsymbol{P}$. Here we consider the case of two inputs and $d$ outputs.
\subsection{No-signaling Polytope $\boldsymbol{P(2,d)}$}
We have two parties, Alice and Bob, who choose from two inputs $X$ and $Y$ $\in \{0,1\}$ and receive outputs $a$ and $b$ with
a joint probability $P(X=a,Y=b)$. We denote the number of distinct outputs associated with inputs X and Y by $d_X^A$ and $d_Y^B$. Any event in this scenario is described as a point in the polytope $\boldsymbol{P(2,d)}$ i.e. the polytope consisting of all no-signaling boxes with two inputs and arbitrary large number of outputs. A vertex of $\boldsymbol{P(2,d)}$ must satisfy (\ref{pos}), (\ref{norm}), (\ref{signal1},\ref{signal2}) and $dim(\boldsymbol{P(2,d)})$ of the positivity inequalities (\ref{pos}) replaced with equalities where
\begin{equation}
\it {dim}(\boldsymbol{P(2,d)})=\displaystyle{\sum\limits_{X,Y=0}^{1}}d_X^A d_Y^B-\displaystyle{\sum\limits_{X=0}^{1}}d_X^A-\displaystyle{\sum\limits_{Y=0}^{1}}d_Y^B
\end{equation}
The extremal points of $\boldsymbol{P(2,d)}$ are of two kinds: partial-output vertices (at least one of the conditions $P(X=a)=0$ or $P(Y=b)=0$ hold) and full-output vertices (all $P(X=a)\neq 0$ and $P(Y=b)\neq 0$)\cite{BLMPPR05}. Partial-output vertices correspond to the vertices of some other polytope $\boldsymbol{\tilde{P}}$ with fewer local dimension (i.e. $d_X^{'A}<d_X^A$ or $d_Y^{'B}<d_Y^B$). On the other hand, the vertices of a polytope $\boldsymbol{\tilde{P}}$, can be extended to vertices of $\boldsymbol{P}$ by assigning a zero probability $P(X=a)=0$ and $P(Y=b)=0$ to extra outcomes. From this mapping it is quite evident that for full-output vertices all outcomes contribute to the no-signaling box. So we need to construct only the full-output vertices for a polytope characterized by $d_X^A$ and $d_Y^B$. The extremal points of the dimension asymmetric cases, where $d_X^A\neq d_Y^B$, will be the full-output extremal points of $d$-outcome polytopes for $d\in\{2,...,min(d_X^A,d_Y^B)\}$.
\subsubsection*{Local Vertices}
Local vertices of polytope $\boldsymbol{P(2,d)}$ correspond to the extremal boxes which realize deterministic strategies using shared randomness only. Allowing for reversible relabeling of the observers' outputs by the local vertices take the following form \cite{BLMPPR05}
\[
 P^{\alpha\beta\gamma\delta}_L =
\begin{cases}
 1 & \text{if } a=\alpha X\oplus\beta, b=\gamma Y\oplus\delta \\
 0 & \text{otherwise}
 \end{cases}
\]\\
where the indices $\alpha,\beta,\gamma,\delta\in\{0,...,\min(d_X^A,d_Y^B)-1\}$ correspond to the reversible relabeling and $\oplus$ denotes sum modulo d, where $d= min(d_X^A,d_Y^B)$.
\subsubsection*{Non-local Vertices}
The non-local vertices of polytope $\boldsymbol{P(2,d)}$ correspond to the strategies which cannot be realized without the observers communicating. Under local reversible relabeling all such non-local vertices take the form \cite{BLMPPR05}
\[
 P^{\alpha\beta\gamma}_{NL} =
 \begin{cases}
 \frac{1}{d} & \text{if } (b\ominus a) =XY\oplus\alpha X\oplus\beta Y\oplus\gamma,\\
 & a,b\in \{1,...,d\}\\
0 & \text{otherwise}
\end{cases}
\]
where the indices $\alpha,\beta,\gamma\in\{0,...,\min(d_X^A,d_Y^B)-1\}$ correspond to the local reversible relabeling, $\oplus$ and $\ominus$ denote sum modulo $d$ and subtraction modulo $d$ respectively, where $d= min(d_X^A,d_Y^B)$.
\section{Results for higher dimensional systems}
\label{sec_res}
\subsection{Results for bipartite $(2,d)$ scenario }\label{res1}
At this point we are ready to present the main results of this work. Let $q_{H}$ be the probability of success of a bipartite two input, $(d_X^A, d_Y^B)$ output conventional Hardy paradox(\ref{hardyOld}) for any non-local vertex of $\boldsymbol{P(2,d)}$. The structure of the conventional Hardy argument(\ref{hardyOld}) suggests that we assign zero probability to all outcomes other than $(1,d_X^A)$ on Alice's side and $(1,d_Y^B)$ on Bob's side, which essentially corresponds to a partial-output vertex of $\boldsymbol{P(2,d)}$. This situation can be mapped to a full-output vertex of $\boldsymbol{P(2,2)}$. Thus the value for $q_{H}$, achieved by any non-local full-output vertex of $\boldsymbol{P(2,2)}$ is
\begin{equation}
q_{H}^{full}= \frac{1}{2}
\end{equation}
and it becomes independent of local dimension. Now moving to the relaxed Hardy argument, let $q_{RH}$ be the probability of success of a bipartite two input, $(d_X^A, d_Y^B)$ output relaxed Hardy paradox(\ref{hardyNew}) for any non-local vertex of $\boldsymbol{P(2,d)}$. It can be easily shown that for a full-output vertex of $\boldsymbol{P(2,d)}$, the maximum number of non-zero elements contributing to the success probability of relaxed Hardy test is $(d-1)$ where $d=min(d_X^A, d_Y^B)$, since these are the only possible events satisfying the following two conditions with the input being $(X_1,Y_1)$
\begin{eqnarray}
(b-a) mod~d = 1;\nonumber\\ 
a,b\in \{1,...,d\} \label{nl}\\
a<b\label{hard}.
\end{eqnarray}
While Eq.(\ref{nl}) refers to the condition for non-zero value of events for a non-local full-output vertex, Eq.(\ref{hard}) denotes the condition for non-zero probability of success of relaxed Hardy's test(\ref{hardyNew}). Thus it can be easily shown that the maximum value of $q_{RH}$ that can be achieved by a full-output vertex of $\boldsymbol{P(2,d)}$ takes the following form
\begin{equation}\label{full2}
q_{RH}^{full}=\frac{d-1}{d}.
\end{equation}
Note that this success probability of relaxed Hardy's test(\ref{full2}) increases with the local dimensions. In the asymptotic limit i.e. for $d=\min(d_X^A, d_Y^B)\rightarrow \infty$, $q_{RH}^{full}$ tends to $1$, which is optimal. Here a natural question is whether these values($q_{H}^{full}$, $q_{RH}^{full}$) are optimal for any finite dimensional correlation in GNST. In the following section we address this question.
\subsection{$q^{opt}$ in GNST} 
\begin{figure}
\includegraphics[scale=0.6]{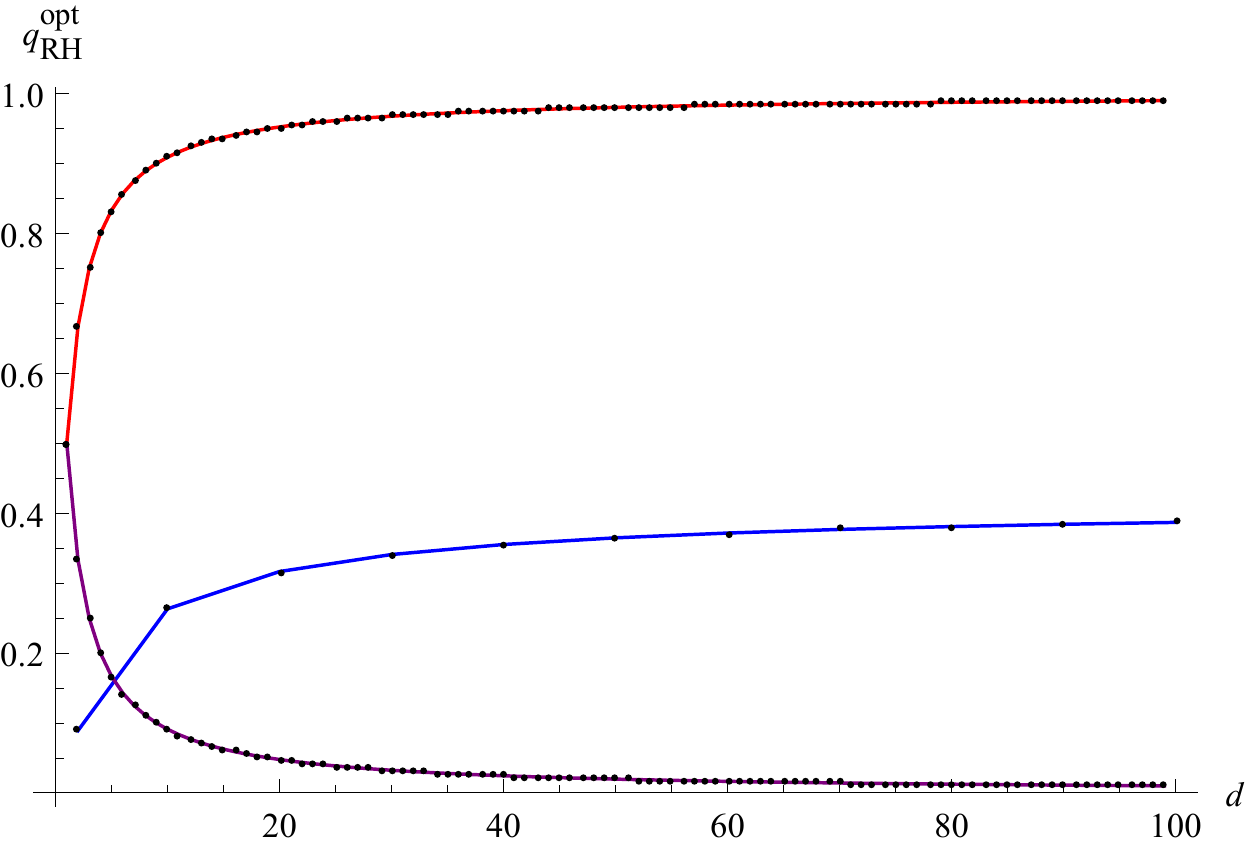}
\caption{(Color on-line). The line in blue shows the increase of $q_{RH}^{opt}$ with increasing system dimension for quantum systems\cite{CCXSWK13}. The red line shows the optimal paradoxical probability $q_{RH}^{opt}$ for generalized no-signaling correlations. The purple line shows the decrease of the contribution of complementary events to $PN$ with increasing local dimension for generalized no-signaling correlations.}
\label{fig1}
\end{figure}

Let us define the no-signaling limit of the probability of success of Hardy's test be $q_H^{opt}$ and relaxed Hardy's test be $q_{RH}^{opt}$, and the correlations that achieve these optimal values be $P_{H}^{opt}$ and $P_{RH}^{opt}$ respectively. Due to the convexity of the no-signaling polytope $\boldsymbol{P(2,2)}$ and $\boldsymbol{P(2,d)}$, it readily follows that $P_{H}^{opt}$ and $P_{RH}^{opt}$ can be written as a probabilistic mixture of local and non-local full-output vertices of $\boldsymbol{P(2,2)}$ and $\boldsymbol{P(2,d)}$ respectively. Thus we can conclude that in any GNST
\begin{eqnarray}
q_{H}^{opt}= \frac{1}{2}\\
q_{RH}^{opt}= \frac{min\{d_X^A,d_Y^B\}-1}{min\{d_X^A,d_Y^B\}}
\end{eqnarray}
Here we see that for an arbitrarily large system dimension the success probability of relaxed Hardy's test(\ref{hardyNew}) tends to its possible maximum (from Eq.\ref{norm},\ref{hardyNew}) value i.e 1 in no-signaling paradigm. This is an interesting feature in tune with the quantum case where the maximum probability of success increases with local dimension\cite{CCXSWK13}. Unlike quantum case considered in \cite{CCXSWK13}, here we consider even the dimension asymmetric scenario($d_X^A\neq d_Y^B$). Fig.(\ref{fig1}) shows the plot for the success probability of relaxed Hardy's test against the dimension of the sub-systems in a generalized no-signaling theory and in quantum theory\cite{CCXSWK13}.

For Hardy paradox with many outcomes we have seen that the paradoxical probability reaches very close to 1 under a generalized no signaling theory. It has great significance. It indicates to the fact that relativistic causality alone does not stop one to demonstrate contradiction with {\it{local realism}} using an argument like (\ref{hardyNew}) with almost $100\%$ success in contrast to the partial success as in quantum case\cite{Har92,CCXSWK13}.
\par
One can interpret this phenomenon by making two consecutive observations-
\begin{observation}
For the generalized Hardy-type argument(\ref{hardyNew}), the optimal $PPC$ corresponding to the optimal $PP$ decreases with increasing minimal dimension d in GNST.  
\end{observation} 

\begin{observation}
$PP\approx PN$ for extremal Non-local correlations with high minimal dimension d in GNST.
\end{observation}

Whereas the first observation indicates that the optimal contribution of the `paradoxical probability' connected with complementary events to the total probability of witnessing Hardy nonlocality decreases with increasing local minimal dimension (purple line in Fig.\ref{fig1}), the second observation tells that in the asymptotic limit the paradoxical probability corresponding to the argument (\ref{hardyNew}) is equal to PN, which is $1$.

\section{Conclusion}
\label{concl}
In comparison to Bell's statistical argument Hardy's paradox is simpler to demonstrate the fact that quantum mechanics contains correlations which can not be simulated with shared randomness alone. Chen {\it et al.}\cite{CCXSWK13} provides the natural generalized version of Hardy's non-locality argument for higher dimensional systems. The authors showed that for $d=2$ it is exactly the old Hardy's non-locality argument. Whereas in the quantum domain for $d>2$, the paradoxical probability of relaxed Hardy's argument increases with $d$. Here we have generalized Chen's conclusion in the no-signaling paradigm. We observe that in any theory that respects relativistic causality, the maximum paradoxical probability of the non-locality argument increases with local dimensions of the two subsystems in bipartite scenario. Finally we conclude our work by providing a proof which emphasizes a simple functional dependence of the paradoxical probability of the generalized Hardy's non-locality argument on local dimensions. This fact indicates that with increasing local minimal dimension, the paradoxical probability corresponding to the relaxed Hardy-type argument approaches the probability of witnessing Hardy nonlocality in GNST. This interpretation of our result also suggests that the non-locality argument proposed in \cite{CCXSWK13} is the most natural higher-dimensional generalization of Hardy's argument\cite{Har92} in two input scenario.

\textbf{Acknowledgment}: The authors thank Prof Guruprasad Kar, Manik Banik and Md. Rejjak Gazi for fruitful discussions. AM acknowledges support from CSIR(Govt. of India) project $09/093(0148)/2012$-EMR-I. RR acknowledges support from the UGC (University Grants Commission, Govt. of India) Start-Up Grant.

\end{document}